\documentclass[12pt,a4]{paper}
\usepackage{a4}

\pdfoutput=1 
\usepackage{minitoc,cite,fancyhdr}
\usepackage{bbm} 
\usepackage{euscript,amssymb,latexsym} 
\usepackage{epsf,epsfig,graphics,subfigure,graphicx,rotating}
\usepackage{amsmath}
\makeatletter
\def\fmslash{\@ifnextchar[{\fmsl@sh}{\fmsl@sh[0mu]}}
\def\fmsl@sh[#1]#2{%
  \mathchoice
    {\@fmsl@sh\displaystyle{#1}{#2}}%
    {\@fmsl@sh\textstyle{#1}{#2}}%
    {\@fmsl@sh\scriptstyle{#1}{#2}}%
    {\@fmsl@sh\scriptscriptstyle{#1}{#2}}}
\def\@fmsl@sh#1#2#3{\m@th\ooalign{$\hfil#1\mkern#2/\hfil$\crcr$#1#3$}}
\makeatother
\global\arraycolsep=2pt 
\def\be{\begin{equation}}
\def\ee{\end{equation}}
\newcommand{\Herwig}{\textsf{Herwig++}}
\newcommand{\fastjet}{\textsf{FastJet}}
\newcommand{\GeV}{\,\mathrm{GeV}}
\newcommand{\TeV}{\,\mathrm{TeV}}
\renewcommand{\d}{\mathrm{d}}

\begin{document}
\begin{flushright}
IPPP/13/17\\
DCPT/13/34\\
MCNET-13-03\\
\end{flushright}
\vspace{16pt}
{\Large \bf Monte Carlo Simulation of Hard Radiation in Decays\\
 in Beyond the Standard Model Physics in \mbox{Herwig++}}\\[0.7cm]
\hspace{-16pt}{\bf\normalsize{Peter Richardson, Alexandra Wilcock}}\vspace{12pt}
\\ 
{\it\normalsize{Institute of Particle Physics Phenomenology, Department of Physics}}
\\ 
{\it\normalsize{University of Durham, DH1 3LE, UK;}}
\\
{\it\normalsize{Email:} }
{\sf\normalsize{peter.richardson@durham.ac.uk, a.h.wilcock@durham.ac.uk}}
\begin{abstract}
We use the POWHEG formalism in the \Herwig\ event generator to match QCD real-emission matrix elements with the parton shower for a range of decays relevant to Beyond the Standard Model physics searches.  Applying this correction affects the shapes of experimental observables and so changes the number of events passing selection criteria.  To validate this approach, we study the impact of the correction on Standard Model top quark decays.  We then illustrate the effect of the correction on Beyond the Standard Model scenarios by considering the invariant-mass distribution of dijets produced in the decay of the lightest Randall-Sundrum graviton and transverse momentum distributions for decays in Supersymmetry.  We consider only the effect of the POWHEG correction on the simulation of the hardest emission in the shower and ignore the normalisation factor required to correct the total widths and branching ratios to next-to-leading order accuracy.

\end{abstract}

\thispagestyle{empty}
\thispagestyle{plain}
\setcounter{page}{1}
\noindent
---------------------------------------------------------------------------------------------------------
\tableofcontents
\noindent
---------------------------------------------------------------------------------------------------------

\clearpage
\section{Introduction} \label{sec:Intro}

For Beyond the Standard Model (BSM) scenarios with additional new particles, the decays of these particles determine the experimental signals we would observe at collider experiments.  If the new particles have a well separated mass spectrum, long decay chains will occur when a heavy new particle is produced.  For decays involving coloured particles, hard quantum chromodynamic (QCD) radiation at each step in the decay chain will alter the structure of the event and therefore the number of events passing experimental selection criteria.  The effects of radiation are also important in models with degenerate new particle mass spectra, where decay chains are typically limited to one step.  Searches for these compressed spectra scenarios look for events in which hard radiation in the initial-state shower recoils against missing transverse energy in the final state to give an observable signal\footnote{See \cite{LeCompte:2011cn} for a recent study.}.  The emission of hard QCD radiation in the decay of new particles could either enhance or reduce this effect and so must be taken into account.  Therefore, accurate simulation of hard radiation in the decays of BSM particles is necessary in order to optimise searches for new physics.  

Monte Carlo event generators use fixed-order matrix elements combined with parton showers and hadronization models to simulate particle collisions.  In the \Herwig\ event generator \cite{Bahr:2008pv, Arnold:2012fq}, the decays of unstable fundamental particles are treated separately from the hard process which produced them, prior to the parton shower phase, using the narrow width approximation.  Decays are generated using the algorithm described in \cite{Richardson:2001df}, which ensures spin correlations are correctly treated.
The parton shower utilises an approximation that resums the leading collinear and leading-colour soft logarithms \cite{Buckley:2011ms} and so does not accurately describe QCD radiation in the regions of phase space where the transverse momenta of the emitted partons are high. The Positive Weight Hardest Emission Generator (POWHEG) formalism \cite{Nason:2004rx} is one method that allows the simulation of high transverse momentum (hard) radiation to be improved upon by using the real-emission matrix element to produce the hardest emission in the shower.  This approach affects both the overall cross sections for inclusive processes and results in local changes to the shapes of distributions sensitive to the hardest emission.  In particular, local changes to observables such as jet transverse momenta are important since they can impact on the proportion of events passing selection criteria in new physics searches.  Since BSM signals often consist of only a few events, this can in turn result in significant changes to the exclusion bounds that can be set. 

The POWHEG formalism has been successfully applied to a wide range of hard production processes, for example \cite{Frixione:2007nw, Alioli:2008gx, Hamilton:2008pd, Alioli:2008tz,  Nason:2009ai, Alioli:2009je, Hamilton:2009za, Re:2010bp, Hamilton:2010mb, Alioli:2010qp, Alioli:2010xa, Platzer:2011bc, Oleari:2011ey, Melia:2011gk, Jager:2011ms, Melia:2011tj, D'Errico:2011sd, D'Errico:2011um, Jager:2012xk, Re:2012zi, Jager:2013mu}, and particle decays \cite{LatundeDada:2008bv, Richardson:2012bn} in the Standard Model (SM) as well as selected BSM processes \cite{Papaefstathiou:2009sr, Bagnaschi:2011tu, Klasen:2012wq, FridmanRojas:2012yh, Jager:2012hd}.  Next-to-leading order (NLO) corrections to BSM particle decays have also previously been studied, for example in \cite{Horsky:2008yi} where the Supersymmetric-QCD correction to the decay \(\tilde{q} \rightarrow q \tilde{\chi} \) was calculated.  In this work, we present results from the implementation of the POWHEG method in \Herwig\ for a range of decays relevant for new physics searches. A similar approach
based on generic spin structures is used to apply a matrix-element correction to hard radiation in particle decays in \textsf{PYTHIA}~\cite{Norrbin:2000uu}.

The POWHEG formalism will be reviewed in Sect.~\ref{sec:POWHEG} and in Sect.~\ref{sec:TopQuark} our implementation of the POWHEG correction will be described in full for the example of top quark decay.  In Sect.~\ref{sec:BSMDecays}, details of the decay modes implemented will be given.  The impact of the correction on the decay of the lightest graviton in the Randall-Sundrum (RS) model \cite{Randall:1999ee} will be studied in Sect.~\ref{sec:RSResults}.  Finally, results from a selection of decays in the Constrained Minimal Supersymmetric Standard Model (CMSSM) will be presented in Sect.~\ref{sec:MSSMResults}.
 
\section{POWHEG Method} \label{sec:POWHEG}
In this section, a brief outline of the POWHEG method is given.  Further details can be found in \cite{Frixione:2007vw}.

In the conventional parton shower approach, the inclusive differential cross section for the highest transverse momentum emission from an \(N\)-body process is given by 
\be
  d \sigma^{\rm{PS}}=B(\Phi_N) d\Phi_N \left[\Delta(p_{T \rm{min}}) \, + \Delta(p_T) d\Phi_R \mathcal{P} \right]. \label{eq:PS}
\ee
Here we are considering a parton shower ordered in terms of the transverse momentum of the emitted parton, \(p_T\).  \( \Phi_N\) are the phase space variables of the \(N\)-body leading-order (LO) process and \(B\) is the Born-level matrix element squared, including the relevant flux factor, such that the total LO cross section is \( \sigma^{\rm{LO}} = \int B(\Phi_N) d\Phi_N\).  \( \mathcal{P}\) is the unregularized Altarelli-Parisi splitting kernel and \( \Phi_R\) is a set of variables parameterizing the phase space of the additional radiated parton.  The radiative phase space is limited to the region \(p_T \left(\Phi_R \right)>p_{T \rm{min}}\), where \(p_{T \rm{min}} \) is a transverse momentum cut-off introduced to regularize the infra-red (IR) divergences in the splitting kernel.  The Sudakov form factor for the parton shower is 
\be
  \Delta(p_T) = \exp \left( - \int d\Phi_R \mathcal{P} \, \Theta \left(p_T \left(\Phi_R\right) -p_T \right) \right). \label{eq:PSSudakov}
\ee
The square bracket in Eq.~\ref{eq:PS} integrates to unity which ensures that the total cross section is given by the LO result.

In the POWHEG approach, the inclusive differential cross section for the hardest emission is given by the QCD NLO differential cross section, that is
\be
  d \sigma^{\rm{PO}}=\bar{B}(\Phi_N) d\Phi_N \left[\Delta^{\rm PO}(p_{T \rm{min}}) + \Delta^{\rm PO}(p_T) d\Phi_R \frac{R(\Phi_N, \Phi_R)}{B(\Phi_N)}\right], \label{eq:PO}
\ee
where \(\bar{B}(\Phi_N) \) is defined by
\begin{align}
  \bar{B}(\Phi_N) = B(\Phi_N) &+ \left[V(\Phi_N) + \int C(\Phi_N, \Phi_R) d\Phi_R\right]  \nonumber \\
&+ \int\left[R(\Phi_N, \Phi_R) d\Phi_R- C(\Phi_N, \Phi_R) d\Phi_R\right]. \label{eq:Bbar}
\end{align}
The real-emission contribution, \(R(\Phi_N, \Phi_R)\), corresponds to the radiation of an additional parton from the LO interaction and the virtual contribution, \(V(\Phi_N)\), comes from the 1-loop correction to the LO process.  \(C(\Phi_N, \Phi_R)\) is a counter term with the same singular behaviour as the real and virtual contributions and is introduced to ensure the two square brackets in Eq.~\ref{eq:Bbar} are separately finite.  The Sudakov form factor appearing in Eq.~\ref{eq:PO} is
\be
  \Delta^{\rm PO}(p_T) = \exp \left( - \int d\Phi_R \frac{R(\Phi_N, \Phi_R)}{B(\Phi_N)}  \, \Theta \left(p_T \left(\Phi_R \right) -p_T \right) \right). \label{eq:POSudakov}
\ee
As with the conventional parton shower approach, the square bracket in Eq.~\ref{eq:PO}  will integrate to unity and hence the total inclusive cross section will be given by the NLO result.

Typically, the counter term, \(C(\Phi_N, \Phi_R)\), can be rewritten as a sum of dipole functions, \(\mathcal{D}_i \), each of which describes the behaviour of the real-emission matrix element in a singular region of phase space, i.e. when the emitted parton becomes soft or collinear to one of the legs in the Born process.  By doing so, the different singular regions can be separated such that Eq.~\ref{eq:POSudakov} becomes a product of Sudakov form factors 
\be
  \Delta^{\rm PO}(p_T) = \prod_i \exp \left( - \int d\Phi_R  \frac{R(\Phi_N, \Phi_R) \mathcal{D}_i} {\sum_j \mathcal{D}_j  B(\Phi_N)}  \, \Theta \left(p_T \left(\Phi_R \right) -p_T \right) \right), \label{eq:POSudakov2}
\ee
each of which describes the non-emission probability in a particular region of phase space specified by the dipole function \(\mathcal{D}_i\). 

When applying the POWHEG method to a parton shower ordered in transverse momentum, the hardest emission is generated first using the POWHEG Sudakov form factor in Eq.~\ref{eq:POSudakov2}.  Subsequent emissions are generated with the normal parton shower Sudakov given in Eq.~\ref{eq:PSSudakov}, with the requirement that no parton shower emission has higher transverse momentum than the emission described by \(R(\Phi_N, \Phi_R)\). However, to allow QCD coherence effects to be included, an angularly ordered parton shower is used in \Herwig.  Ordering the parton shower in terms of an angular variable means the first emission in the shower may not be the hardest.  The POWHEG approach can be reconciled with angularly ordered parton showers by dividing the shower into several steps \cite{Nason:2004rx}.  The hardest emission in the shower is generated first using the POWHEG Sudakov form factor and the value of the angular evolution variable corresponding to this emission is determined.  An angularly ordered shower, running from the shower starting scale down to the scale of the hardest emission, is then generated.  This \textit{truncated} parton shower simulates coherent soft wide-angle radiation. The hardest emission is then inserted and the shower continues until the IR cut-off of the evolution variable is reached.  Finally, in both stages of the parton shower, emissions generated by the shower are discarded if they have higher transverse momentum than the emission generated using the POWHEG Sudakov form factor.

\section{Top Quark Decays} \label{sec:TopQuark}
In this section, we describe our implementation of the POWHEG formalism for the example of a top quark decaying to a \(W\) boson and a bottom quark.  To implement the full POWHEG correction to this decay, the Born configuration must be generated according to Eq.~\ref{eq:Bbar} and the hardest emission in the parton shower simulated using Eq.~\ref{eq:POSudakov2}.  However, in this work we consider only the effect of the POWHEG correction on the simulation of the hardest emission in the shower and hence generate the Born configuration using only \(B \left(\Phi_N \right)\), the leading order contribution in Eq.~\ref{eq:Bbar}.  As such, we use the existing LO \Herwig\ implementation of top quark decay and modify the shower such that the hardest emission is generated according to Eq.~\ref{eq:POSudakov2}.  Justification for excluding the normalisation factor of the POWHEG correction will be given in Sect.~\ref{sec:BSMDecays}.

Application of the POWHEG correction to top quark decays, along with top quark pair production in \(e^+ e^-\) collisions, has been previously studied in \cite{LatundeDada:2008bv} for massless bottom quarks.  In this work, we retain the mass of the bottom quark throughout.  

\subsection{Implementation in \Herwig\ }\label{sec:TopQuarkImp}
In \Herwig, the decays of fundamental particles are performed in the rest frame of the decaying particle.  In this frame, we are free to choose the orientation of the \(W\) boson to be along the negative \(z\)-direction and so, at LO, the bottom quark is orientated along positive \(z\)-direction.  The squared, spin and colour averaged matrix element for the LO process is given by
\be
 |\mathcal{M}_B|^2 = \frac{g^2} {4m_w^2} \left(m_t^4+m_b^4-2m_w^4+m_t^2m_w^2+m_b^2m_w^2-2m_t^2m_b^2 \right),
\ee
where \(m_t\), \(m_b\) and \(m_W\) are the masses of the top quark, bottom quark and \(W\) boson respectively and \(g\) is the weak interaction coupling constant. The relevant CKM factor has been set equal to 1.

The squared, spin and colour averaged matrix element for the \(\mathcal{O} \left(\alpha _s \right)\) real-emission correction to the decay \(t \rightarrow W b\) is
\begin{multline}
|\mathcal{M}_R|^2 = g^2 g_s^2 C_F \left\{ - \frac{|\mathcal{M}_B|^2} {g^2} \left(\frac{p_b}{p_b.p_g} - \frac{p_t} {p_t.p_g} \right)^2 + \right. \\ \left. \left(\frac{p_g.p_t}{p_b.p_g} + \frac{p_b.p_g}{p_t.p_g} \right) \left(1 + \frac{m_t^2}{2m_w^2} + \frac{m_b^2}{2m_w^2} \right) - \frac{1}{m_w^2} \left(m_t^2+m_b^2 \right)\right\},
\end{multline}
where \(g_s\) is the strong coupling constant, \(C_F= \frac{4}{3}\) and \(p_t\), \(p_b\), \(p_W\) and \(p_g\) are the four-momenta of the top quark, bottom quark, \(W\) boson and gluon.  In general, the orientation of the decay products in the three-body final state is such that the emitting parton absorbs the transverse recoil coming from the emission of the gluon and the spectator particle continues to lie along the negative \(z\)-direction.  When the radiation originates from the top quark, the bottom quark effectively acts as the emitting particle so that we remain in the rest frame of the top quark.  Therefore, for emission from both the top and the bottom quarks, the momenta of the decay products are
\be
p_W = \left(E_W,0,0,-\sqrt{E_W^2 - m_W^2} \right),
\ee
\be
p_b = \left(E_b, -p_T \cos \left( \phi \right), -p_T \sin \left( \phi \right), \sqrt{E_b^2 - p_T^2 -m_b^2}  \right),
\ee
\be
p_g = \left(E_g, p_T \cos \left( \phi \right), p_T \sin \left( \phi \right), \sqrt{E_g^2 - p_T^2 }  \right),
\ee
where \(E_x\) is the energy of particle \(x\), and \(p_T\) and \(\phi\) are the transverse momentum and azimuthal angle of the gluon.

The Lorentz invariant phase space element, \(\d \Phi_R\), describing the emission of the additional gluon is obtained from the relation
\be
 \d \Phi_3 = \d \Phi_2 \d \Phi_R ,
\ee
where
\be
\d \Phi_N = \left(2 \pi \right)^4 \delta^4 \left(p_t - \sum_{i=1}^N p_i \right) \prod_{i=1}^{N} {\frac{\d ^3\mathbf{p_i}}{2E_i(2\pi)^3}}
\ee
and \(\mathbf{p_i}\) is the three-momentum of particle \(i\).  We choose to parameterize the radiative phase space in terms of the transverse momentum, \(p_T\), rapidity, \(y\), and azimuthal angle, \(\phi\), of the gluon and so find
\be
 \d \Phi_R = J \d p_T dy \d \phi,
\ee
where the Jacobian factor, \(J\), is \footnote{\(\lambda(x,y,z) = \sqrt{x^2+y^2+z^2-2xy-2xy-2yz} \).}
\be
J = \frac{1}{8\pi^2} \frac{m^2_t p_T |\mathbf{p_W}|^2}{\lambda(m_t^2, m_W^2, m_b^2) [|\mathbf{p_W}|(m_t-p_T\cosh y)-E_W p_T\sinh y]}.
\ee
This parametrization has the advantage of simplifying the Heaviside function in the POWHEG Sudakov form factor to a lower limit in the integration over \(p_T\).

The final components required for the implementation of the POWHEG Sudakov form factor in Eq.~\ref{eq:POSudakov2} are the dipole functions, \(\mathcal{D}_i\), which describe the singular behaviour of the real-emission matrix element.  We use the dipole functions defined in the Catani-Seymour subtraction scheme, details of which can be found in \cite{Catani:1996vz, Catani:2002hc}, to describe the singular behaviour resulting from emissions from the decay products.  The dipole used to describe radiation from the top quark is as follows
\be
\mathcal{D}_i = \frac{-4 \pi C_F \alpha_s } {E^2_g} |\mathcal{M}_B|^2. \label{eq:initialDipole}
\ee
It contains only soft enhancements since, in the top quark rest frame, collinear enhancements are suppressed.

Using the above information, the hardest emission in the shower can then be generated according to Eq.~\ref{eq:POSudakov2} using the veto algorithm\footnote{A good description of the veto algorithm can be found in~\cite{Sjostrand:2006za}.}, which proceeds as follows:
\begin{enumerate}
\item Trial values of the radiative phase space variables are generated.  The transverse momentum of the emission is generated by solving
\be
 \Delta^{\rm{over}}\left(p_T \right) = \exp \left(- \int^{p^{\rm{max}}_T}_{p_T} \frac{C\left(y_{\rm{max}}-y_{\rm{min}}\right)}{p_T\left( \Phi_R\right)} \d p_T\left( \Phi_R\right) \right) = \mathcal{R} , \label{eq:SudakovOver}
\ee
where \(p^{\rm{max}}_T = \frac{\left(m_t -m_W \right)^2 -m_b^2} {2 \left(m_t - m_W \right)}\) is the maximum possible \(p_T\) of the gluon.  \(y_{\rm{max}}\) and \(y_{\rm{min}}\) are the upper and lower bounds on the gluon rapidity, chosen to overestimate the true rapidity range.  \(C\) is a constant chosen such that the integrand in Eq.~\ref{eq:SudakovOver} always exceeds the integrand in Eq.~\ref{eq:POSudakov2} and \(\mathcal{R}\) is a random number distributed uniformly in the range \([0,1]\).  Values of \(y\) and \(\phi\) are generated uniformly in the ranges \([y_{\rm{min}}, y_{\rm{max}}]\) and \([0, 2 \pi]\) respectively;
\item If \(p_T < p^{\rm{min}}_T\), no radiation is generated and the event is hadronized directly.  We set \(p^{\rm{min}}_T=1\GeV \) throughout this work; 
\item If \(p_T \geq p^{\rm{min}}_T\), the momenta of the \(W\) boson, bottom quark and gluon are calculated using the generated values of the radiative variables.  Doing so, yields two possible values of \(E_W\) that must both be retained and used in the remainder of the calculation.  If the resulting momenta do not lie within the physically allowed region of phase space, we veto this configuration, set \(p^{\rm{max}}_T = p_T\) and return to step 1; 
\item Events within the physical phase space are accepted with a probability given by the ratio of the true to overestimated integrands in Eqs.~\ref{eq:POSudakov2} and~\ref{eq:SudakovOver} respectively.  If the event is rejected, we set \(p^{\rm{max}}_T = p_T\) and return to step 1;
\end{enumerate}
Using this procedure, a trial emission is generated for each dipole, \(\mathcal{D}_i\), in Eq.~\ref{eq:POSudakov2}.  The configuration which gives the highest \(p_T\) emission is selected.  The existing \Herwig\ framework, detailed in~\cite{Hamilton:2008pd}, is then used to generate the remainder of the parton shower.

\subsection{Parton Level Results}\label{sec:TopQuarkRes}

\begin{figure}[t]
\begin{subfigure}
    \centering
    \begin{sideways}
      \includegraphics[trim=24mm 49mm 24mm 10mm, clip, scale=0.34]{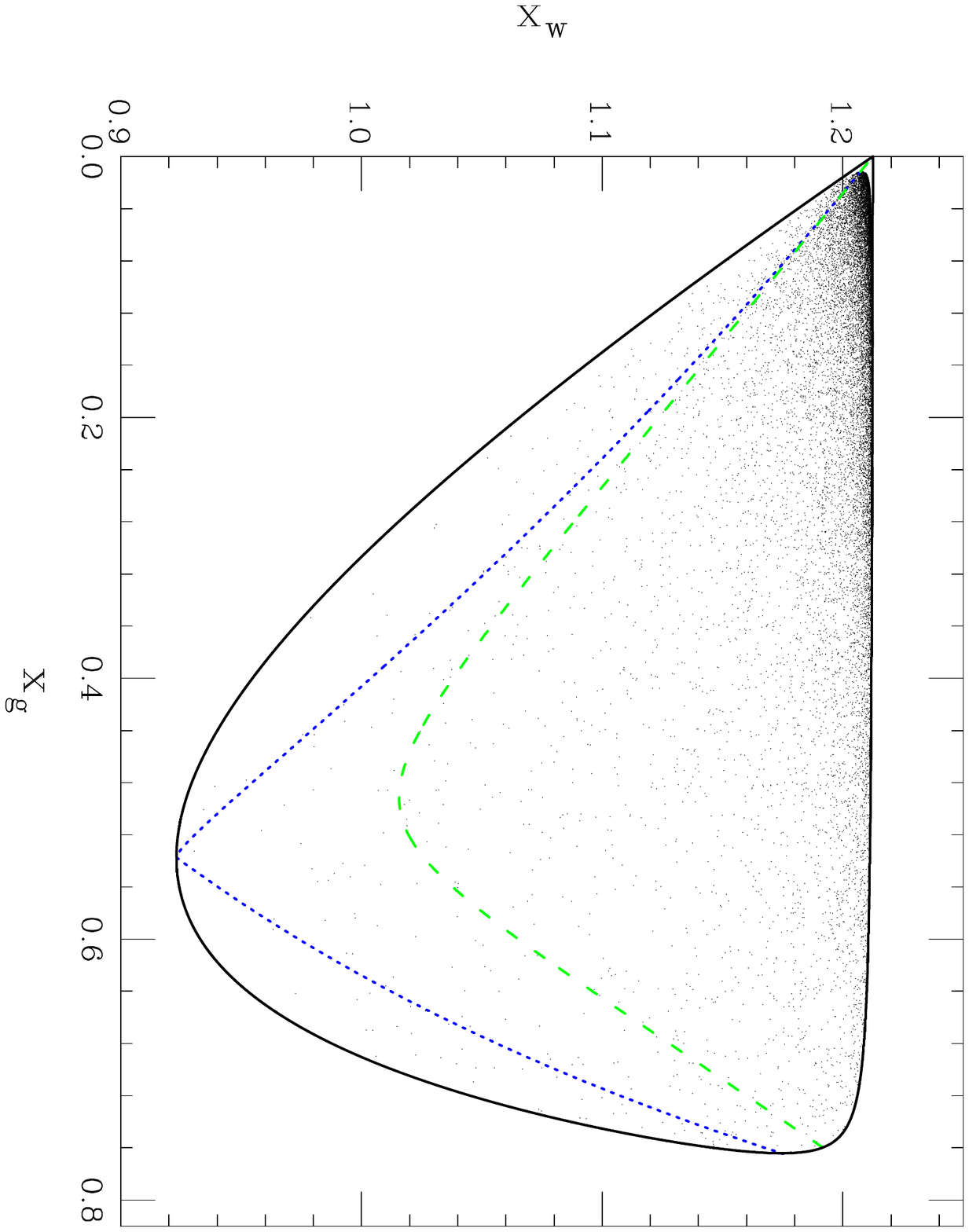}
    \end{sideways}
  \end{subfigure}
  \begin{subfigure}
    \centering
    \begin{sideways}
    \includegraphics[trim=24mm 49mm 24mm 10mm, clip, scale=0.34]{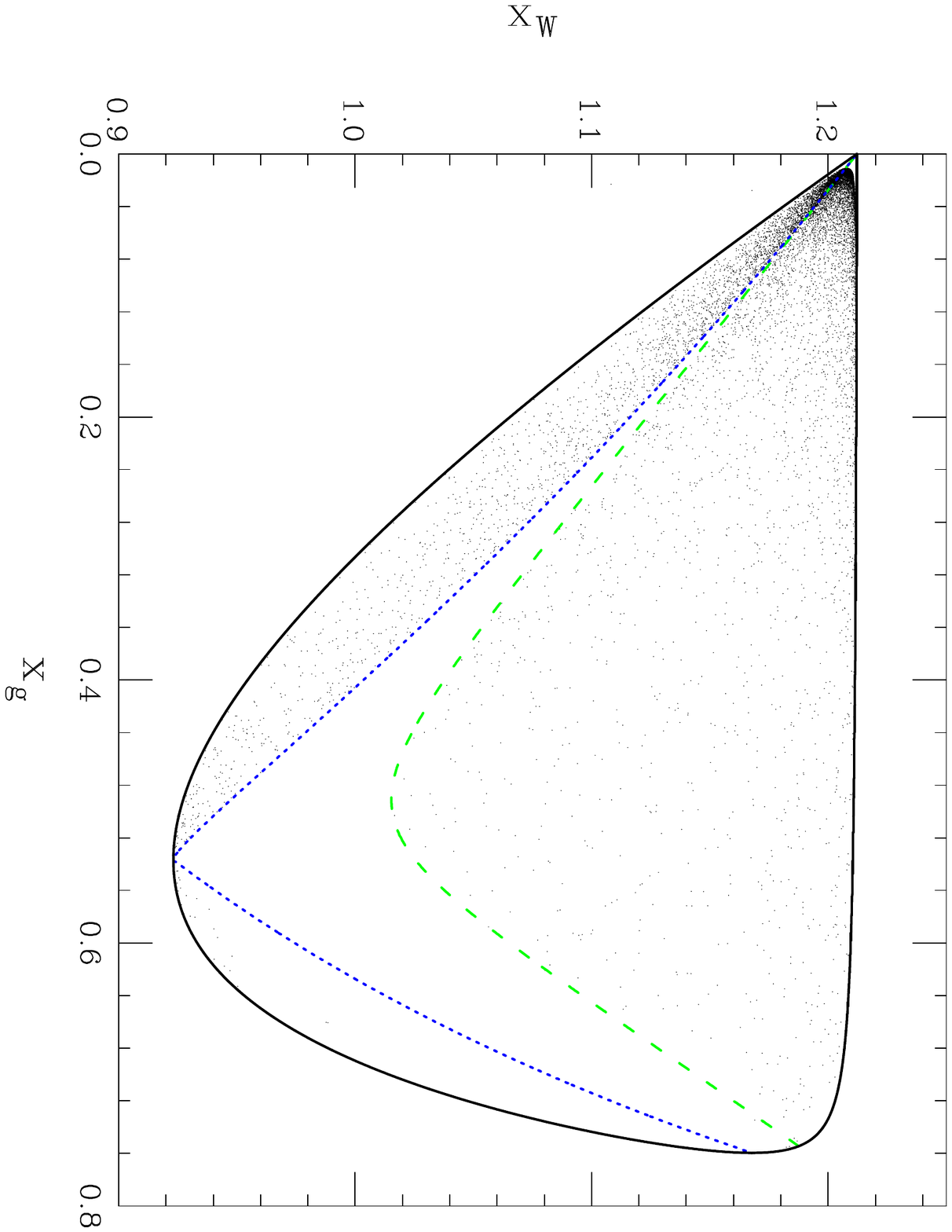}
    \end{sideways}
  \end{subfigure}
  \caption{Dalitz distributions for the decay \(t \rightarrow W b\) with (left) and without (right) the POWHEG style correction.  The black outline indicates the physically allowed region of phase space.  In the conventional parton shower approach, the region above the green dashed line is populated with radiation from the bottom quark and the regions below the blue dotted lines with radiation from the top quark.  These boundaries correspond to the limits of the parton shower with symmetric phase space partitioning.}
  \label{fig:top_Dalitz}

\end{figure}

To validate our implementation of the algorithm described in Sec.~\ref{sec:TopQuarkImp}, Dalitz style plots were generated for the decay \(t \rightarrow W b\) and are shown in Fig.~\ref{fig:top_Dalitz}.  The Dalitz variables, \(x_W\) and \(x_g\), were defined by the relation, \(x_i = \frac{2E_i}{m_t}\), where \(E_i\) is the energy of particle \(i\) in the rest frame of the top quark.  The left-hand plot in Fig.~\ref{fig:top_Dalitz} shows the distribution obtained using the POWHEG style correction.  In this case, \(x_g\) is the energy fraction of the gluon generated using the full real-emission matrix element.  The distribution on the right-hand side of Fig.~\ref{fig:top_Dalitz} was generated using the conventional parton shower, limited to one emission in the final state, and so here \(x_g\) is the energy fraction of a gluon produced using the parton shower splitting kernels.  On both distributions, the black outline indicates the physical phase space boundaries.  The enclosed area is divided into a section populated by radiation from the bottom quark (above the green dashed line), sections populated by radiation from the top quark (below the blue dotted lines) and a \textit{dead region} (between the blue dotted and green dashed lines) that corresponds to hard gluon radiation and is not populated by the conventional parton shower.  These boundaries correspond to the theoretical limits of the \Herwig\ parton shower with symmetric phase space partitioning, described in~\cite{Gieseke:2003rz}, in which the starting values of the shower evolution variables for the top and bottom quarks are chosen such that the volumes of phase space accessible to emissions from each quark are approximately equal. 

As expected, in both plots we see a high density of points in the limit \(x_g \rightarrow 0\), corresponding to soft gluon emission.  The POWHEG corrected distribution also has a concentration of points along the upper physical phase space boundary where \(x_W\) is maximal and emissions are collinear to the bottom quark.  The density of points along the upper boundary is reduced in the parton shower distribution and points are instead concentrated along the lower boundary of the bottom quark emission region.  As discussed in~\cite{Gieseke:2003rz}, the parton shower approximation agrees with the exact matrix element in the case of collinear radiation from the bottom quark but overestimates it elsewhere in the bottom quark emission region.  The factor by which the parton shower approximation exceeds the exact matrix element, increases towards the lower boundary of the region and therefore we see an excess of points near the boundary.  The parton shower distribution also has a high density of points in the top quark emission region for \(x_g \lesssim 0.53\).  This enhancement is again the result of the parton shower approximation overestimating the exact matrix element in this area~\cite{Gieseke:2003rz}.  In general, we see that the parton shower in \Herwig\ produces areas of high emission density which do not correspond to physically enhanced areas of phase space and therefore has a tendency to overpopulate hard regions of phase space.  On the other hand, the POWHEG emission is distributed according to the exact real-emission matrix element and so correctly populates the physically enhanced regions of phase space with no additional spurious high density regions.  Finally, we also see that the POWHEG corrected distribution fills the dead region of phase space that is not populated by the standalone parton shower.

\begin{figure}[t]
  \centering
  \begin{subfigure}
    \centering
    \includegraphics[trim=0mm 0mm 00mm 00mm, clip, scale=0.65]{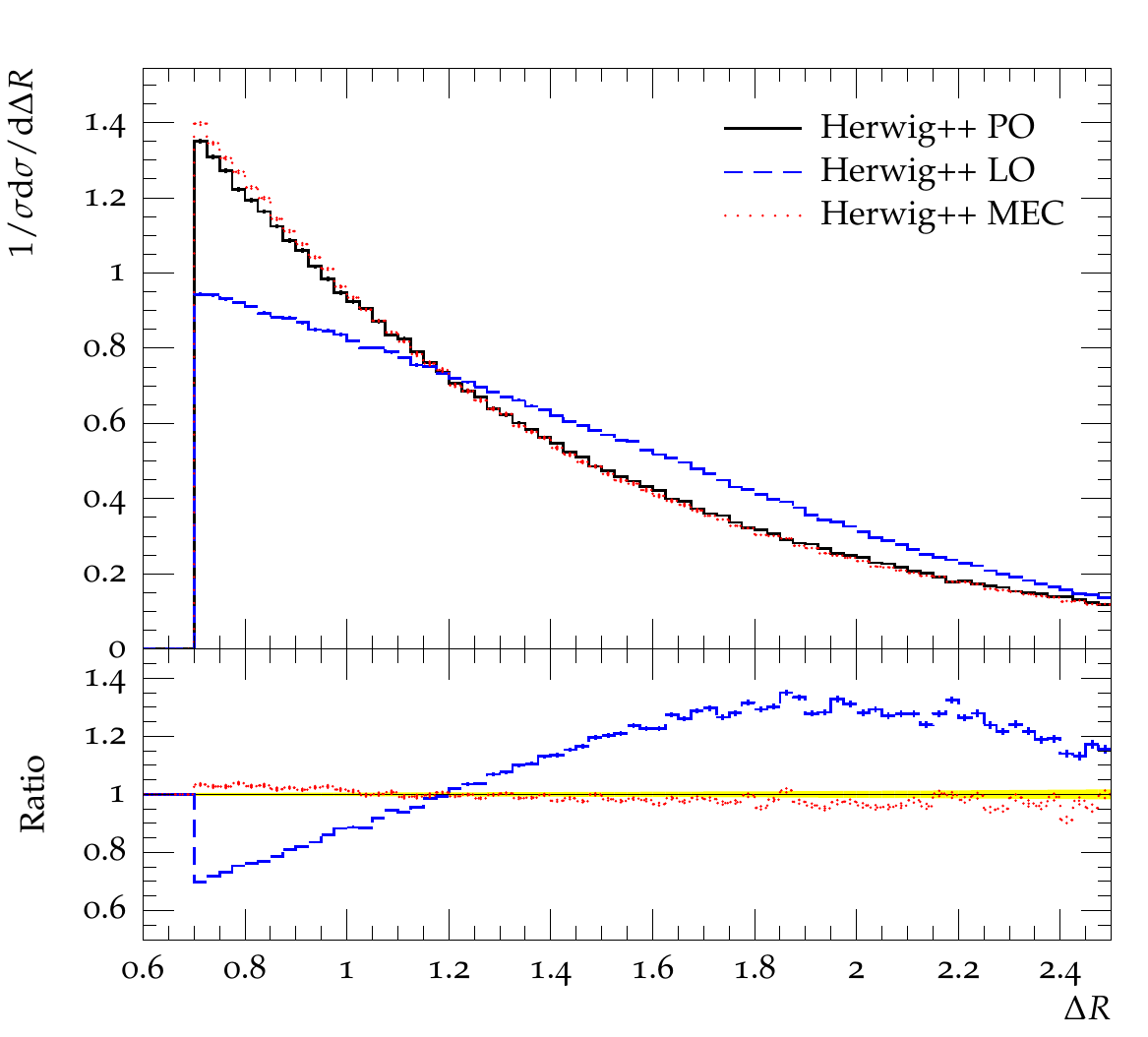}
  \end{subfigure}
  \begin{subfigure}
    \centering
    \includegraphics[trim=0mm 0mm 0mm 00mm, clip, scale=0.65]{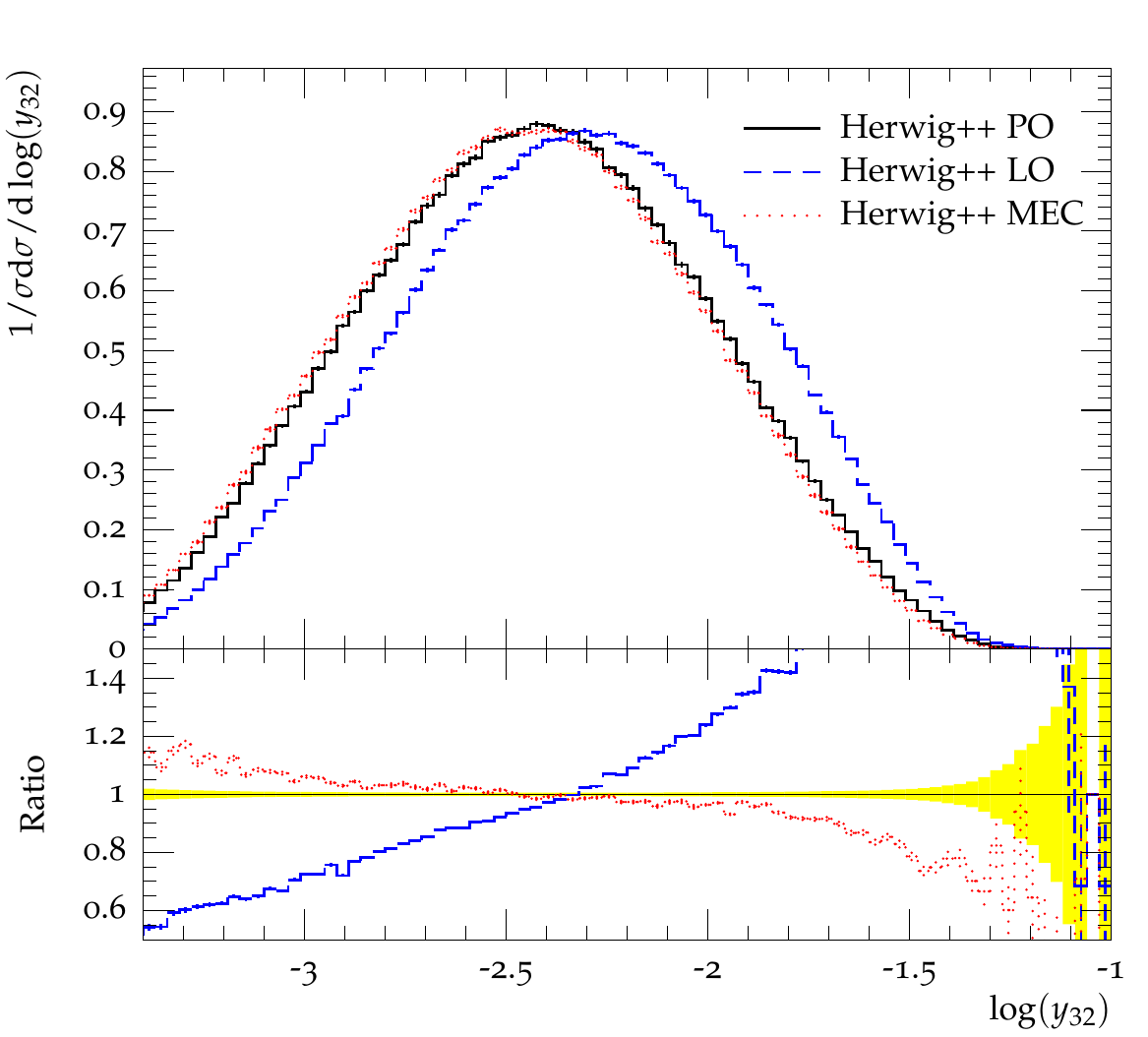} 
  \end{subfigure}
  \caption{Comparison of distributions generated using the standalone parton shower with those generated using a matrix element or POWHEG style correction to the decay \(t \rightarrow W b\).  Parton level \(e^+e^- \rightarrow t \bar{t}\) events were generated at \(\sqrt{s}=360 \GeV\).  The left-hand plot shows the distribution of the minimum jet separation, \(\Delta R\), and the right-hand plot the logarithm of the jet measure, \(y_{32}\).  }
  \label{fig:top_parton}

\end{figure}

To study the impact of the POWHEG style correction on top quark decays, parton level \(e^+ e^- \rightarrow t \bar{t}\) events were simulated and analysed as in~\cite{Hamilton:2006ms}. Events were generated at a centre-of-mass energy close to the \(t \bar{t}\) threshold, \(\sqrt{s} =360 \GeV\), to minimize the effects of radiation from the initial-state shower.  Unless otherwise stated, in this study we use the default set of tuned perturbative and non-perturbative parameters, or \textit{event tune}, in \Herwig\ version 2.6 \cite{Arnold:2012fq}.  Final-state partons were clustered into three jets using the \fastjet \cite{Cacciari:2011ma} implementation of the \(k_T\) algorithm.  The \(W\) bosons were decayed leptonically and their decay products excluded from the jet clustering.  Events were discarded if they contained a jet with \(p_{T}<10 \GeV\) or the minimum jet separation\footnote{\(\Delta R = \min_{ij}\sqrt{\Delta \eta_{ij} ^2 + \Delta \phi_{ij} ^2}\) where the indices \(i,j\) run over the three hardest jets and \mbox{\(i \neq j\)}.  \(\Delta \eta_{ij}\) and \(\Delta \phi_{ij}\) are the differences in pseudorapidity and azimuthal angle of jets \(i\) and \(j\) respectively.}, \(\Delta R\) , did not satisfy \( \Delta R \geq 0.7\).  Using events that passed these selection criteria, differential distributions were plotted of  \( \Delta R \) and \(\log \left(y_{32}\right)\), where \(y_{32}\) is the value of the jet resolution parameter\footnote{\(y_{32} = \frac{2} {s} \min_{ij} \left(\min \left(E_i^2, E_j^2 \right) \left(1-\cos \theta_{ij} \right)\right)\) where again the indices \(i,j\) run over the three hardest jets with \(i \neq j\). \(E_i\) is the energy of jet \(i\) and \(\theta_{ij}\) the polar angle between jets \(i\) and \(j\).} at which a three jet event is classified as a two jet event.  The resulting distributions are shown in the left and right-hand plots in Fig.~\ref{fig:top_parton}.  Distributions generated using the normal parton shower and the parton shower including the POWHEG style correction, are shown by the blue dashed and black solid lines respectively.  The red dotted lines in Fig.~\ref{fig:top_parton} show the distributions obtained when the existing implementation of hard and soft matrix element corrections (MEC)~\cite{Hamilton:2006ms} are applied to the normal \Herwig\ parton shower.  Hard matrix element corrections use the full \(t \rightarrow W b g\) matrix element to distribute emissions in the dead regions of phase space that are not populated by the parton shower.  Soft matrix element corrections use the full real-emission matrix element to correct emissions generated by the parton shower that lie outside the areas of phase space where the parton shower approximation is valid, i.e. away from the soft and collinear limits.  Applying these corrections ensures that the hardest emission in the shower is generated according to the exact matrix element, therefore, we expect a high level of agreement between the POWHEG and matrix element corrected distributions.  The bottom panel in each plot shows the ratio of the parton shower and matrix element corrected distributions to the POWHEG corrected distribution.  In both plots, we include error bars indicating the statistical uncertainty.

As discussed in~\cite{Hamilton:2006ms}, applying the matrix element corrections has the effect of softening both the \(\Delta R \) and \(\log \left(y_{32} \right) \) distributions.  This is due to the soft matrix element correction rejecting a portion of the high \(p_T\) emissions generated by the parton shower.  The magnitude of the observed effect illustrates the importance of matching the parton shower to the exact matrix element in high \(p_T\) regions.  As expected, the distributions generated using the POWHEG style and matrix element corrections are very similar although, for both variables, the POWHEG style correction yields slightly harder distributions.  The discrepancies between the distributions are the result of a number of subtle differences between the POWHEG and matrix element correction schemes.  Firstly in the matrix element correction approach, events in the dead region are generated using the fixed-order real-emission matrix element only, without any Sudakov suppression, and subsequent showering of the resulting configuration is simulated starting from the \(1 \rightarrow 3 \) process.  However, in the POWHEG approach the hardest emission in the shower is reinterpreted such that the conventional parton shower instead begins from the Born hard configuration.  The scale of the hardest emission is generated, and then the shower proceeds as normal except that the hardest emission is fixed at the generated scale.  In addition to this, the soft matrix element correction is applied to all emissions in the parton shower which are the hardest so far.  Normally this leads to the correction of both the hardest emission and a number of other emissions with large values of the evolution parameter, but smaller transverse momentum.  These differences all contribute to the discrepancies between the POWHEG style and matrix element corrected distributions although it is unclear which would have the largest effect.  However, the difference between the POWHEG style and matrix element corrected results is comparatively small.  The agreement between the two approaches serves to further validate our implementation of the POWHEG formalism.  Finally, we note that the POWHEG style approach is preferable to the original matrix element correction scheme since it is significantly simpler to implement in \Herwig.


\section{Decays of BSM Particles} \label{sec:BSMDecays}
As discussed in Sect.~\ref{sec:Intro}, it is important that the simulation of QCD radiation in the decays of BSM particles is done in the most accurate way possible.  In this work, we present results illustrating the effect of consistently matching the QCD real-emission matrix element with the parton shower in \Herwig\ through the POWHEG formalism.  This technique has been applied to a range of decays that occur in most of the well studied BSM scenarios.  Tab.~\ref{tab:spins} shows the combinations of incoming and outgoing spins for which this method is used and each spin structure is implemented for the colour flows given in Tab.~\ref{tab:colour}. However, models with coloured tensor particles are beyond the scope of this work and therefore decays involving incoming tensor particles were limited to colour flows in which the tensor is a colour singlet.

The LO and real-emission matrix elements appearing in the POWHEG Sudakov form factor in Eq.~\ref{eq:POSudakov} are calculated using helicity amplitude methods to correctly incorporate spin correlations \cite{Richardson:2001df}.  The dipole functions, \(\mathcal{D}_i \), are defined as in the Catani-Seymour dipole subtraction method\cite{Catani:1996vz, Catani:2002hc} when describing radiation from the decay products.  In this approach, dipoles describing quasi-collinear radiation from massive vector bosons are not well defined.  Therefore, the Fermion-Fermion-Vector, Scalar-Scalar-Vector and Tensor-Vector-Vector decays are limited to the situation where any final-state coloured vector particles are massless.  The Vector-Fermion-Fermion and Vector-Scalar-Scalar decays do, however, include radiation from  massive incoming vector particles.  Decays are performed in the rest frame of the decaying particle \cite{Bahr:2008pv} and therefore the dipole describing the singular behaviour of this particle will only contain a universal soft contribution.  This is a well defined, spin-independent function given, for the example colour flow \(3 \rightarrow 3\,0\), by Eq.~\ref{eq:initialDipole}.

Finally, in this work we focus solely on the effect of the POWHEG correction on the simulation of the hardest emission in the shower and have not implemented the normalisation factor coming from the presence of \(\bar{B}\) rather than \(B\) in Eq.~\ref{eq:PO}.  In many cases, the partial widths and branching ratios used in the simulation are calculated by an external program, for example SDECAY \cite{Muhlleitner:2003vg}, and so already include NLO corrections.  These values are then passed to \Herwig\ by means of a spectrum file in SUSY Les Houches Accord format \cite{Allanach:2004ub,Allanach:2008qq}.  In cases where the calculation of the widths and branching ratios is performed in \Herwig, generated distributions can be rescaled by a global normalisation factor to achieve NLO accuracy for suitably inclusive observables when the necessary calculations exist.

\renewcommand{\arraystretch}{1.05}
\begin{table}
\parbox{.45\linewidth}{
\begin{center}
\small
\begin{tabular}{|c|c|}
  \hline
  \textbf{Incoming} & \textbf{Outgoing} \\   \hline
  Scalar & Scalar Scalar \\ \hline
  Scalar & Scalar Vector* \\ \hline
  Scalar & Fermion Fermion \\ \hline
  Fermion & Fermion Scalar \\ \hline
  Fermion & Fermion Vector* \\ \hline
  Vector & Scalar Scalar \\ \hline
  Vector & Fermion Fermion \\ \hline
  Tensor & Fermion Fermion \\ \hline
  Tensor & Vector Vector* \\ \hline
\end{tabular}
\end{center}
\caption{Spin combinations for which the POWHEG correction has been applied.  Corrections to the decays marked * are not included for massive, coloured vector particles. }
\label{tab:spins}
}
\hfill
\parbox{.45\linewidth}{
\renewcommand{\arraystretch}{1.325}
\begin{center}
\small
\begin{tabular}{|c|c|}
  \hline
  \textbf{Incoming} & \textbf{Outgoing} \\ \hline
  0 & 3 \(\bar{3}^{\dag}\) \\ \hline
  0 & 8 \(8^{ \dag}\) \\ \hline
  3 & 3 0 \\ \hline
  \(\bar{3}\) & \(\bar{3}\) 0 \\ \hline
  3 & 3 8 \\ \hline
  \(\bar{3}\) & \(\bar{3}\) 8 \\ \hline
  8 & 3 \(\bar{3}\) \\ \hline
\end{tabular}
\end{center}
\caption{Colour flows for which the POWHEG correction has been applied. For tensor particles, corrections are only included for colour flows marked \(^{\dag}\).}
\label{tab:colour}
}
\end{table}

\section{Results} 
\subsection{Randall-Sundrum Graviton}\label{sec:RSResults}
The effect of applying the POWHEG correction to the decay of the lightest RS graviton was investigated using the \Herwig\ implementation of the RS model.  LHC proton-proton collisions with a centre-of-mass (CM) energy of \(\sqrt{s}=8 \TeV\) were simulated.   The lightest graviton, \(G\), was produced as a resonance and allowed to decay via \(G \rightarrow gg \) and \(G \rightarrow q \bar{q}\) for \(q = u,d,s,c,b \).  The mass of the graviton was chosen to be \(m_G = 2.23 \TeV\) which corresponds to the lower bound on the allowed graviton mass for the coupling \(k/ \bar{M}_{pl} = 0.1 \) in \cite{Aad:2012cy}.  An analysis based on the ATLAS experiment's search for new phenomena in dijet distributions \cite{ATLAS:2012pu} was then carried out.  Jets were constructed using the \fastjet \cite{Cacciari:2011ma} implementation of the anti-\(k_t\) algorithm \cite{Cacciari:2008gp} with the energy recombination scheme and a distance parameter \(R=0.6\).  Jets  with \(|y| \geq 4.4 \) were discarded, where \(y\) is the rapidity of the jet in the \(pp\) CM frame.  Events with less than two jets passing this constraint were vetoed.  The rapidities of the two highest \(p_T\) jets in the \(pp\) CM frame are given by \(y_{1}\) and \(y_{2}\).  In the dijet CM frame formed by the two hardest jets, their corresponding rapidities are \(y_{*}\) and \(-y_{*}\) where \(y_{*}= \frac{1}{2} (y_1 - y_2)\).  Events not satisfying \(|y_{*}| < 0.6\) and  \(|y_{1,2}| < 2.8\) were discarded.  The dijet invariant mass, \(m_{jj}\), was formed from the vector sum of the two hardest jet momenta and events were vetoed if \(m_{jj} \leq 1.0 \TeV\).

\begin{figure}[t]
  \centering
  \begin{subfigure}
    \centering
    \includegraphics[trim=0mm 0mm 00mm 00mm, clip, scale=0.65]{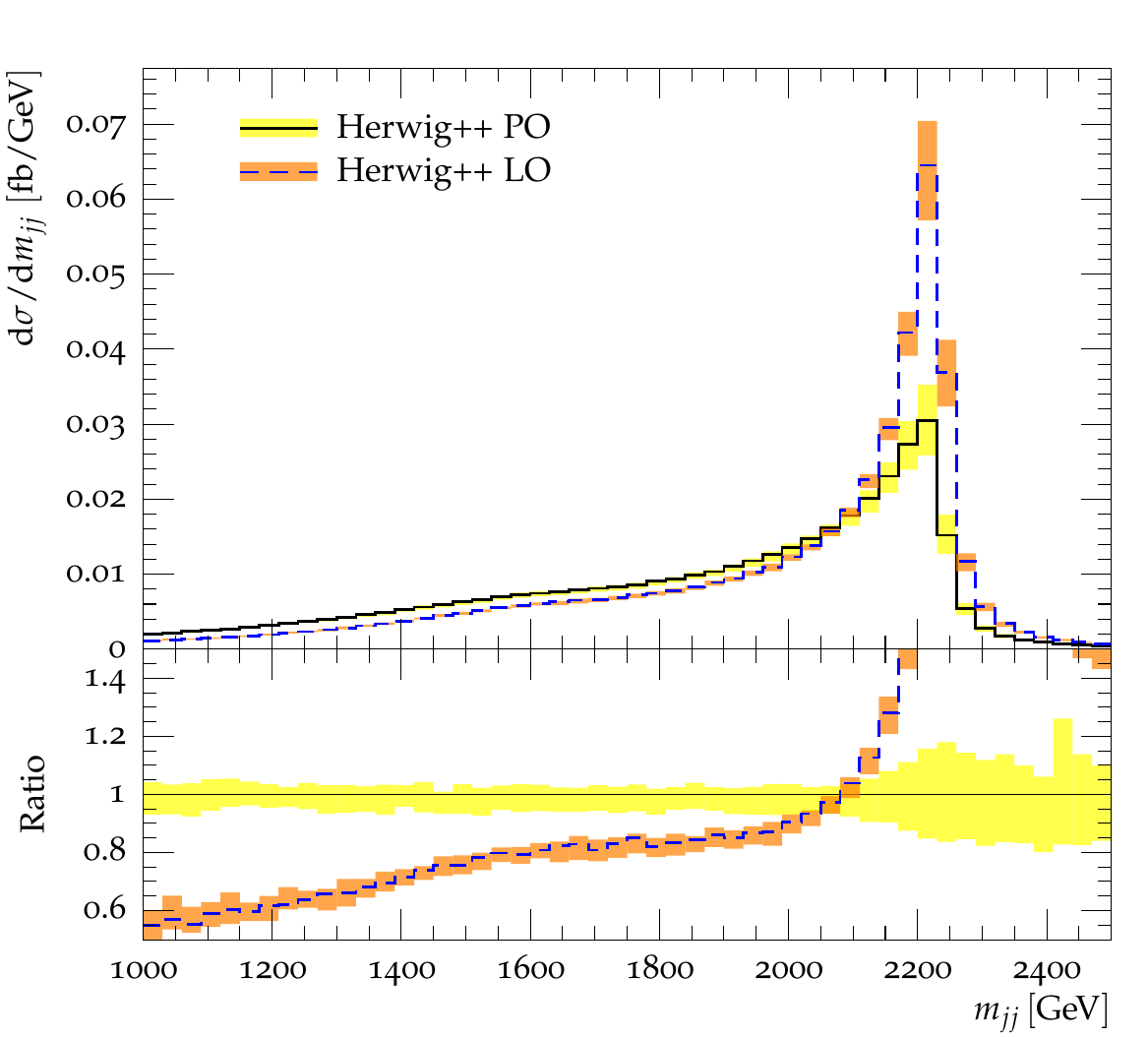}
  \end{subfigure}
  \begin{subfigure}
    \centering
    \includegraphics[trim=0mm 0mm 00mm 00mm, clip, scale=0.65]{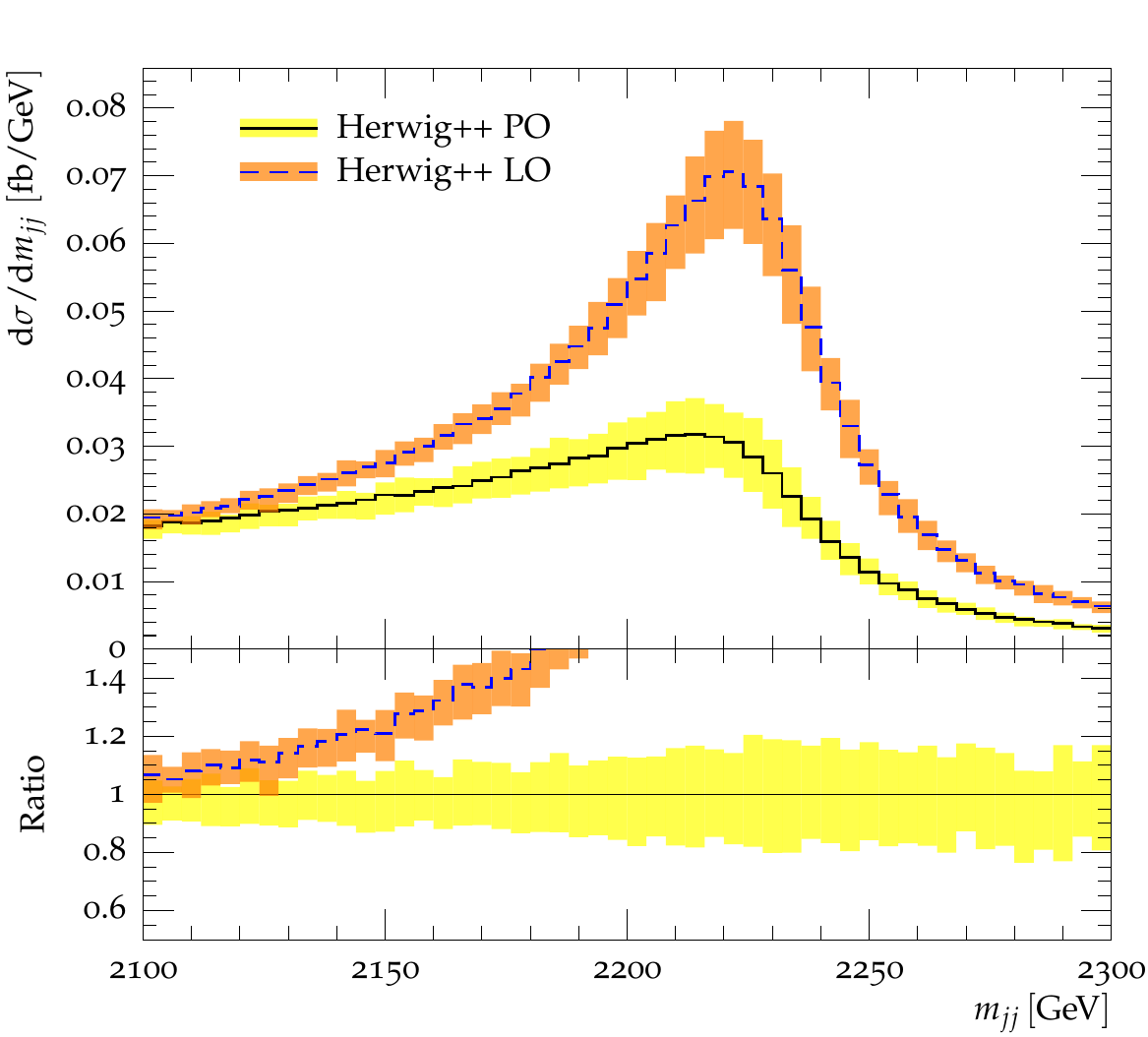}  
  \end{subfigure}
  \caption{Dijet invariant mass distribution for the lightest RS graviton decaying to jets.  The left-hand plot shows the distribution in the full range \mbox{\(1.0 \TeV \leq m_{jj} \leq 2.5 \TeV \)} while the right-hand plot emphasises the effect on the peak region \(2.1 \TeV \leq m_{jj} \leq 2.3 \TeV \).  The mass of the graviton was \(m_G = 2.23 \TeV\) and the coupling \(k/ \bar{M}_{pl} = 0.1 \).  LHC events were simulated with \(\sqrt{s}=8 \TeV\).  The yellow and orange bands were generated by varying the event tune parameters in the POWHEG corrected and conventional parton shower distributions respectively.}
  \label{fig:mass distribution}

\end{figure}

The dijet mass distribution after the above selection criteria were applied, is shown in the left-hand plot in Fig.~\ref{fig:mass distribution}.  The blue dashed line shows the invariant mass distribution for the LO matrix element combined with the parton shower while the black solid line shows the result including the POWHEG correction to the graviton decay.  Both distributions were generated using the optimum set of tuned perturbative and non-perturbative parameters found in\cite{Richardson:2012bn}.  From Fig.~\ref{fig:mass distribution}, we see that including the POWHEG correction causes a decrease of \(\mathcal{O} \left( 40 \% \right) \) in the number of events in the region \(2.1 \TeV \leq m_{jj} \leq 2.3 \TeV \).  This effect is highlighted in the right-hand plot in Fig.~\ref{fig:mass distribution}, which shows the dijet mass distribution in this range.  In the conventional parton shower approach, the majority of the graviton's momentum will be carried by the two partonic decay products.  When the POWHEG correction is applied, the highest \(p_T\) emission in the shower will typically be quite hard and so a significant fraction of the the graviton's momentum will be missed by considering the invariant mass of only the hardest two jets, therefore shifting the distribution to lower values of \(m_{jj}\).  

To give an estimate of the uncertainty arising from our choice of event tune, the dijet mass distributions were generated at ten points in the event tune parameter space and error bands were created showing the maximum and minimum values from the resulting set of distributions.  A description of the varied parameters can be found in\cite{Richardson:2012bn} and their values at each of the ten points are given in Tab.\(\,\)2 of\cite{Richardson:2012bn}.  The error bands are shown in yellow and orange for the distributions with and without the POWHEG correction respectively.  The impact of the POWHEG correction is still clearly evident once this uncertainty has been taken into account.

\subsection{Constrained Minimal Supersymmetric Standard Model} \label{sec:MSSMResults}
In addition to the results presented in Sect.~\ref{sec:RSResults}, the effect of the POWHEG correction was also studied in the context of the CMSSM model.  The high scale parameters of the model were chosen to be \(m_0 = 1220 \GeV\), \(m_{1/2} = 630 \GeV\), \(\tan{\beta} = 10\), \(A_0 = 0\) and \(\mu >0\).  This point lies just outside the exclusion limits set by the ATLAS experiment in \cite{ATLAS:2012ona}.  The corresponding weak scale parameters and decay modes were calculated using ISAJET 7.80 \cite{Paige:2003mg} and the resulting masses of the Supersymmetric (SUSY) particles relevant to this study are given in Tab.~\ref{tab:mass}. The \Herwig\ implementation of the MSSM model was used to generate LHC \(pp\) collisions at a centre-of-mass energy of \(\sqrt{s}=8 \TeV\).  Here we focus on the effect of the correction to the parton shower and so hadronization and the underlying event are not simulated.  In the following sections, the impact of the POWHEG correction on two archetypal decays is presented.  In both cases, the decaying SUSY particle is pair produced in the hard process and the two subsequent decays are then analysed separately in the rest frame of the decaying particle.  Dalitz style distributions were produced, as described in Sec.~\ref{sec:TopQuarkRes}, for both the POWHEG corrected emission and the normal parton shower limited to one final-state emission.  In addition, transverse momentum distributions of the hardest jet not coming from a visible decay product were also studied.  To do so, the full parton shower was generated, with and without the POWHEG style correction, and events were analysed by clustering all visible final-state particles into jets using the \fastjet\ implementation of the anti-\(k_T\) algorithm with the energy recombination scheme and \(R=0.4\).  Jets with \(p_T \leq 20 \GeV\) or \(|\eta|>4.0 \) were discarded.  Events were required to have at least \(n+1\) jets passing the selection criteria, where \(n\) is the number of visible decay products.     

\begin{table}
\centering
\begin{tabular}{|c|c|c|c|}
  \hline
  \(m_{\tilde{u}_L}\) & \(m_{\tilde{g}}\) & \(m_{\tilde{t}_1}\) & \(m_{\tilde{\chi}_1^0}\) \\ \hline
  \(1812.91 \GeV\ \) & \(1546.56 \GeV\ \) & \(1278.14 \GeV\ \) & \(279.22 \GeV\ \) \\ \hline
\end{tabular}
\caption{Masses of the SUSY particles relevant to the decays studied in Secs.~\ref{sec:chiResults} and~\ref{sec:gluinoResults}.  Values were obtained using ISAJET 7.80 with the high scale parameters  \(m_0 = 1220 \GeV\), \(m_{1/2} = 630 \GeV\), \(\tan{\beta} = 10\), \(A_0 = 0\) and \(\mu >0\). }
\label{tab:mass}
\end{table}

\subsubsection{\(\tilde{u}_L \rightarrow u \, \tilde{\chi}_1^0\)} \label{sec:chiResults}

\begin{figure}[t]
\begin{subfigure}
    \centering
    \begin{sideways}
      \includegraphics[trim=24mm 49mm 24mm 10mm, clip, scale=0.34]{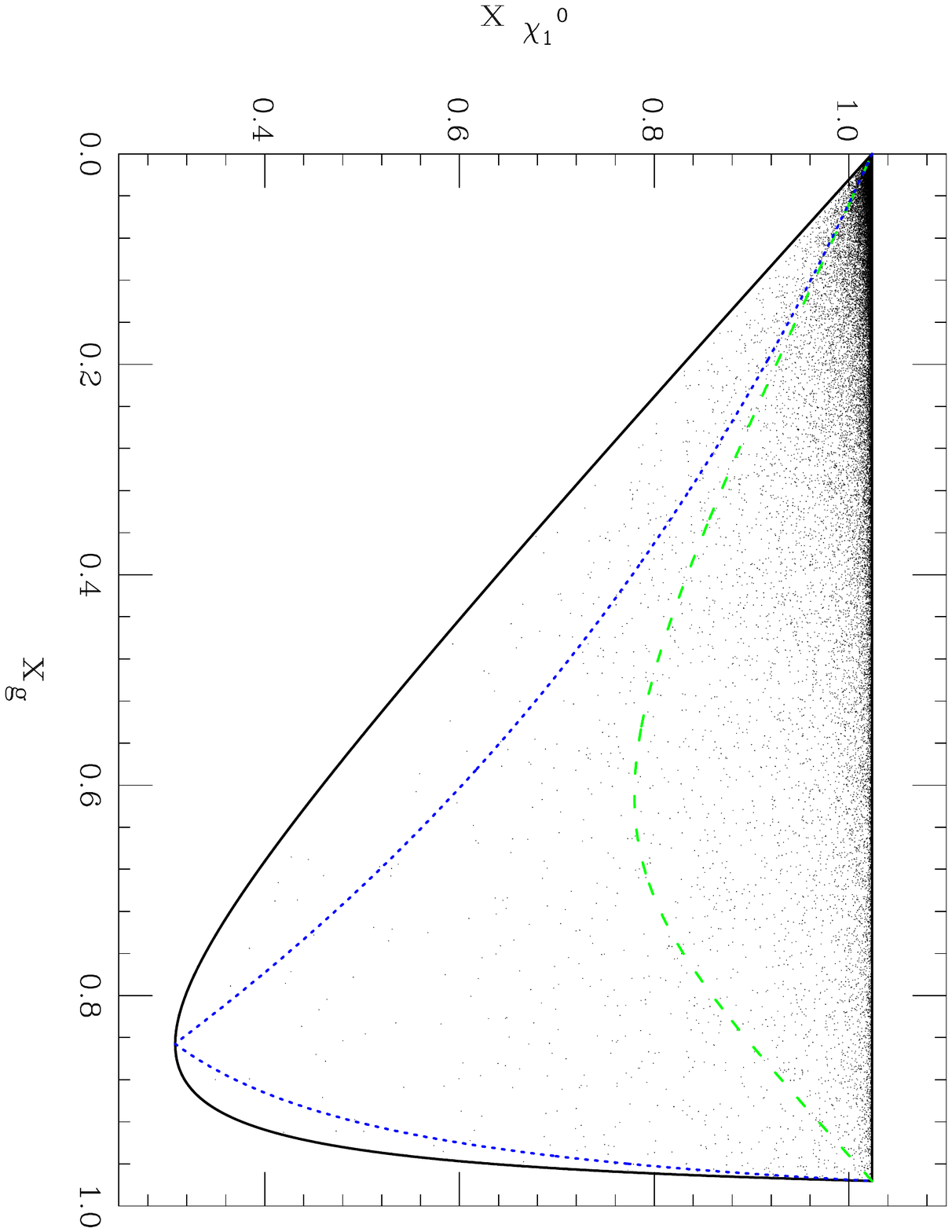}
    \end{sideways}
  \end{subfigure}
  \begin{subfigure}
    \centering
    \begin{sideways}
    \includegraphics[trim=24mm 49mm 24mm 10mm, clip, scale=0.34]{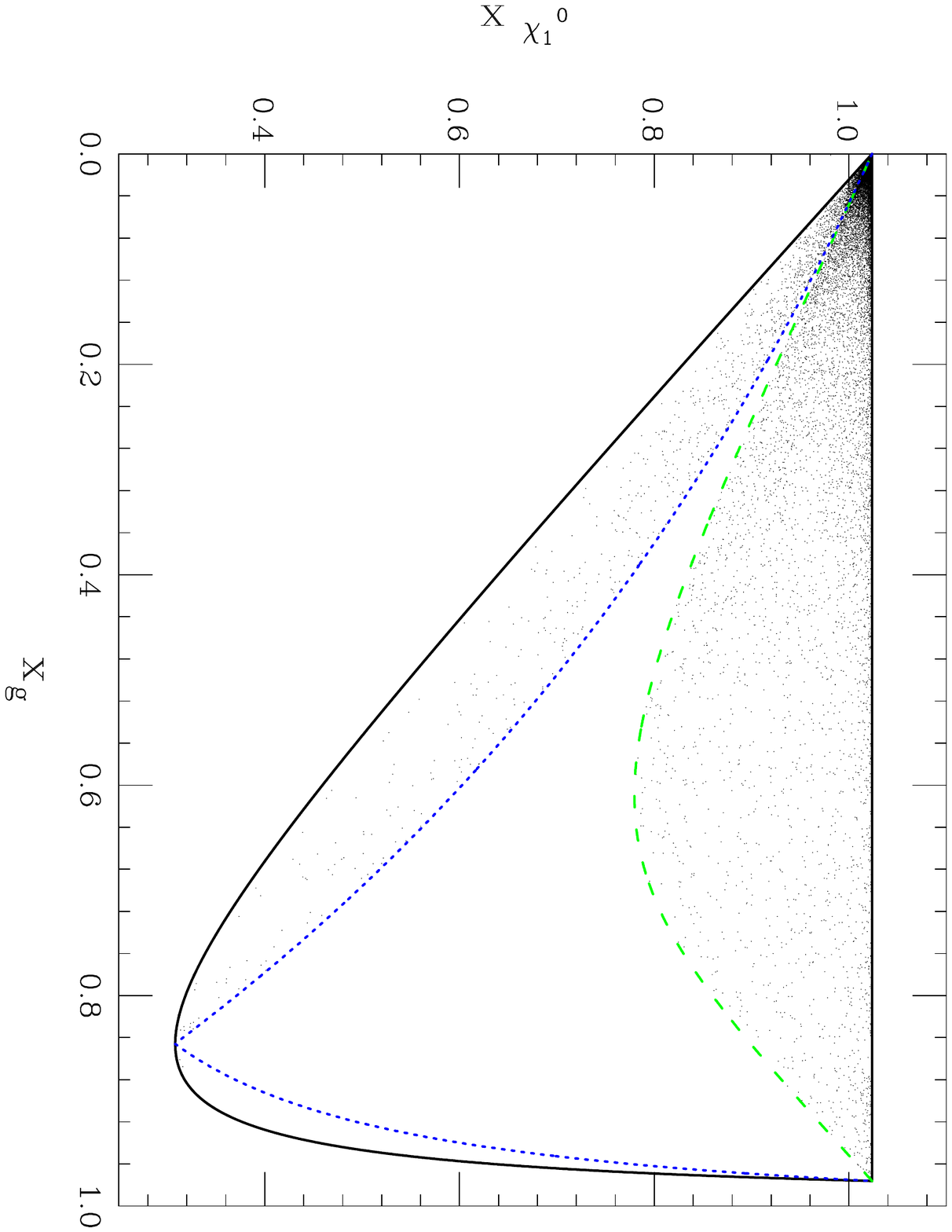}
    \end{sideways}
  \end{subfigure}
  \caption{Dalitz distributions for the decay \(\tilde{u}_L \rightarrow u \, \tilde{\chi}_1^0\) with (left) and without (right) the POWHEG style correction.  The black outline indicates the physically allowed region phase space.  In the conventional parton shower approach, the region above the green dashed line is populated with radiation from the up quark and the regions below the blue dotted lines with radiation from the \(\tilde{u}_L\).  These boundaries correspond to the limits of the parton shower with symmetric phase space partitioning.}
  \label{fig:squark_Dalitz}

\end{figure}

Events were generated in which \(\tilde{u}_L\) and its associated anti-particle were produced and then decayed via the mode \(\tilde{u}_L \rightarrow u \, \tilde{\chi}_1^0\).  Dalitz style distributions with and without the POWHEG correction were produced and are shown in the left and right-hand plots in Fig.~\ref{fig:squark_Dalitz}.  The black outline indicates the kinematic limits of phase space and the green dashed and blue dotted lines are the boundaries of the emission regions of the conventional parton shower with the most symmetric choice of shower phase space partitioning.  Emissions from the up quark populate the area above the green dashed line, while the regions below the blue dotted lines are filled by emissions from the \(\tilde{u}_L\).  The area between the green and blue lines is the dead zone, unpopulated by the normal parton shower.  In the POWHEG corrected distribution, points are concentrated in the soft region as \(x_g \rightarrow 0\) and along the upper boundary of physical phase space where the gluon in collinear to the up quark.  However, in the normal parton shower distribution fewer points lie along the upper physical phase space boundary and instead there is an concentration of points in the \(\tilde{u}_L\) emission region with \( x_g \lesssim 0.85\) and along the lower boundary of the up quark emission region.  In analogy to the case of top quark decay, it is likely that these unphysical high density regions are due to the parton shower kernels overestimating the exact real-emission matrix element.  Finally, we see that including the POWHEG correction ensures that the region of phase space inaccessible to the normal parton shower is populated.

\begin{figure}[h!]
  \centering
  \includegraphics[trim=0mm 0mm 00mm 00mm, clip, scale=0.6]{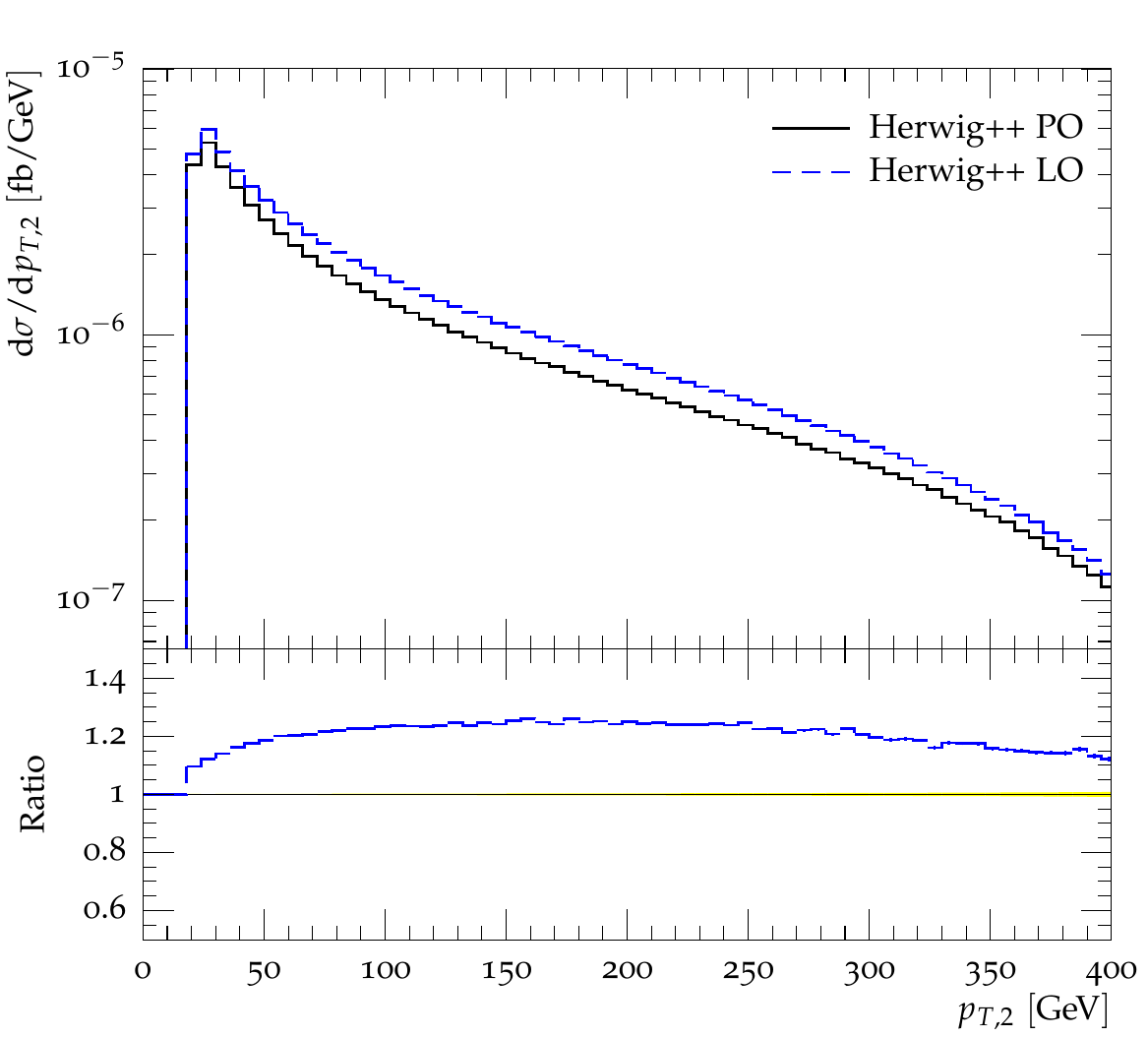}
  \caption{Transverse momentum distributions of the the second hardest jet in the decay \(\tilde{u}_L \rightarrow u \, \tilde{\chi}_1^0\) in the rest frame of the \(\tilde{u}_L\).  Events were generated with and without the POWHEG correction using the CMSSM model with  \mbox{\(m_0 = 1220 \GeV\)}, \(m_{1/2} = 660 \GeV\), \(\tan{\beta} = 10\), \(A_0 = 0\) and \(\mu >0\) at the LHC with \(\sqrt{s}=8 \TeV\). }
  \label{fig:squark distribution}

\end{figure}

Differential distributions of the transverse momentum of the subleading jet\footnote{Jets are ordered in terms of their transverse momentum such that \(p_{T,1} > p_{T,2} > p_{T,3} \) etc.}, \(p_{T,2}\), in each decay were also generated and are shown in Fig.~\ref{fig:squark distribution}.  The blue dashed line corresponds to the distribution generated using the LO matrix element combined with the parton shower while the black solid line shows the result with the POWHEG correction to the decay applied.  The bottom panel in Fig.~\ref{fig:squark distribution} shows the ratio of the parton shower and POWHEG corrected results and in both distributions error bars are included to indicate statistical uncertainty.  As demonstrated in Fig.~\ref{fig:squark_Dalitz}, the parton shower has a tendency to over-populate the hard regions of phase space.  Hence, including the POWHEG correction reduces the \(p_T\) of the hardest emission in the decay.  This phenomenon is reflected in the \(p_{T,2}\) distributions.  When the POWHEG correction is applied, the \(p_{T,2}\) distribution is softened such that there is a reduction in the number of events passing the jet \(p_T\) selection criteria of \(\mathcal{O} \left(20 \% \right)\).  The softening is less pronounced at low values of \(p_{T,2}\) where the parton shower splitting kernels give a good approximation to the exact matrix element.  Here the standalone parton shower and POWHEG corrected distributions are similar.  At larger values of \(p_{T,2}\), the impact of the POWHEG correction is again reduced as, in this region, the subleading jet in the POWHEG corrected distribution typically has a significant contribution from partons generated by the normal parton shower in addition to the hardest emission coming from the POWHEG correction.

\subsubsection{\(\tilde{g} \rightarrow \tilde{t}_1 \,  \bar{t}\)} \label{sec:gluinoResults}

\begin{figure}[t]
\begin{subfigure}
    \centering
    \begin{sideways}
      \includegraphics[trim=24mm 49mm 24mm 10mm, clip, scale=0.34]{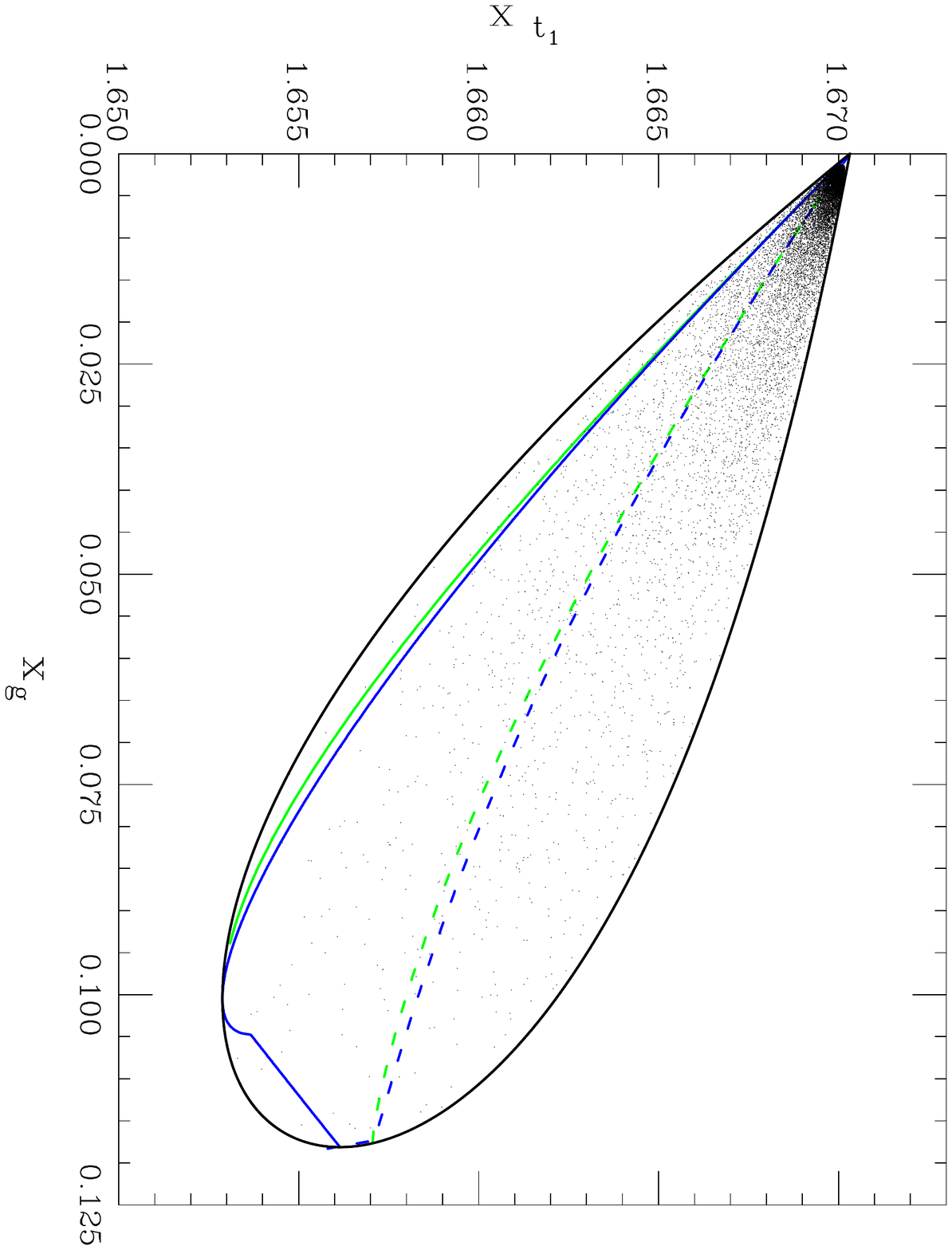}
    \end{sideways}
  \end{subfigure}
  \begin{subfigure}
    \centering
    \begin{sideways}
      \includegraphics[trim=24mm 49mm 24mm 10mm, clip, scale=0.34]{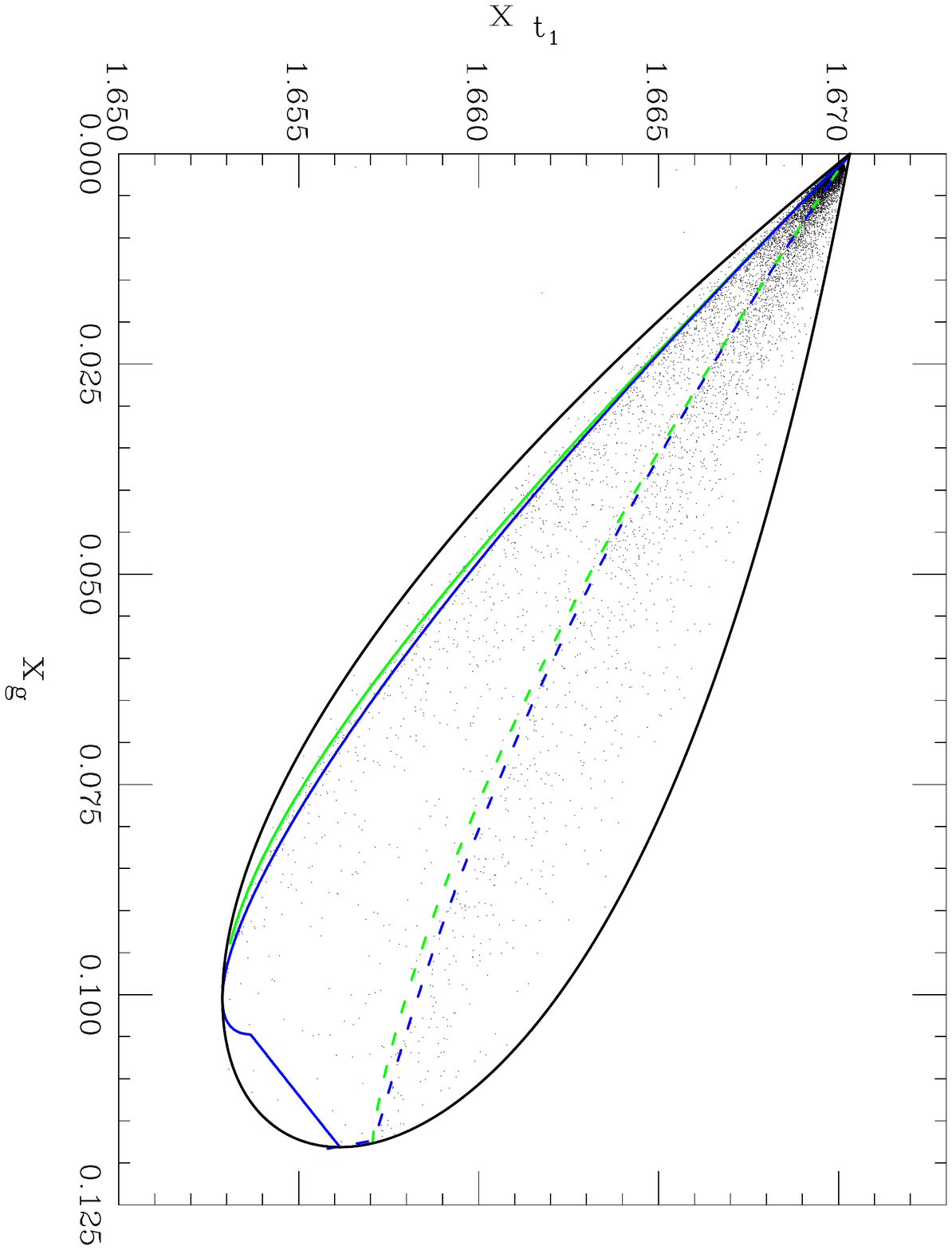}
    \end{sideways}
  \end{subfigure}
  \caption{Dalitz distributions for the decay  \(\tilde{g} \rightarrow  \tilde{t}_1 \, \bar{t}\) with (left) and without (right) the POWHEG style correction applied.  The solid (dashed) coloured lines indicate the parton shower emission regions when the \(\bar{t}\)  \(\left(\tilde{t}_1\right)\) absorbs the transverse recoil of the emission.  The solid (dashed) pale green line shows the lower (upper) boundary for radiation from the \(\bar{t}\) \(\left(\tilde{t}_1 \right) \).  The dark blue solid (dashed) lines are the equivalent upper (lower) boundaries for radiation from the \(\tilde{g}\).  All boundaries correspond to the case of symmetric phase space partitioning and the black outline shows the kinematically allowed region of phase space.}
  \label{fig:gluino_Dalitz}

\end{figure}

Finally, we investigate the impact of the POWHEG style correction on the decay mode \(\tilde{g} \rightarrow \tilde{t}_1 \, \bar{t}\).  The left and right-hand plots in Fig.~\ref{fig:gluino_Dalitz} show Dalitz distributions for this decay with and without the POWHEG correction respectively.  In both plots, the black outline indicates the kinematically allowed region phase space.  The solid coloured lines show the boundaries of the parton shower emission regions in the scenario where the \(\bar{t}\) absorbs the \(p_T\) of the gluon and the \(\tilde{t}_1\) is orientated along the negative \(z\)-axis in the \(\tilde{g}\) rest frame.  The region above the pale green line is populated by emissions from the \(\bar{t}\) and the areas below the dark blue lines are filled by emissions from the \(\tilde{g}\).  In this scenario, the two emission regions overlap and there is no region of phase space left unpopulated by the parton shower.  The dashed coloured lines indicate the emission boundaries of the parton shower when the \(\tilde{t}_1\) absorbs the transverse recoil of the emission and the \(\bar{t}\) is aligned with the negative \(z\)-axis.  The pale green dashed line is the upper limit for emissions coming from the \(\tilde{t}_1\) and the dark blue dashed lines are the lower boundaries from emissions from the \(\tilde{g}\).  From the left-hand plot of Fig.~\ref{fig:gluino_Dalitz}, we see that the majority of points in the POWHEG corrected distribution are concentrated in the soft region of phase space.  High density regions corresponding to emissions collinear to the \(\bar{t}\) or \(\tilde{t}_1\) are suppressed due to the large masses of the decay products.  In the parton shower distribution, points are concentrated in the soft region and along the lower boundary of the \(\bar{t}\) and dashed \(\tilde{g}\) emission regions.  The latter two unphysical regions of over-population again highlight the importance of correcting hard emissions in the parton shower using the exact real-emission matrix element.  

\begin{figure}[h!]
  \centering
  \includegraphics[trim=0mm 0mm 00mm 00mm, clip, scale=0.6]{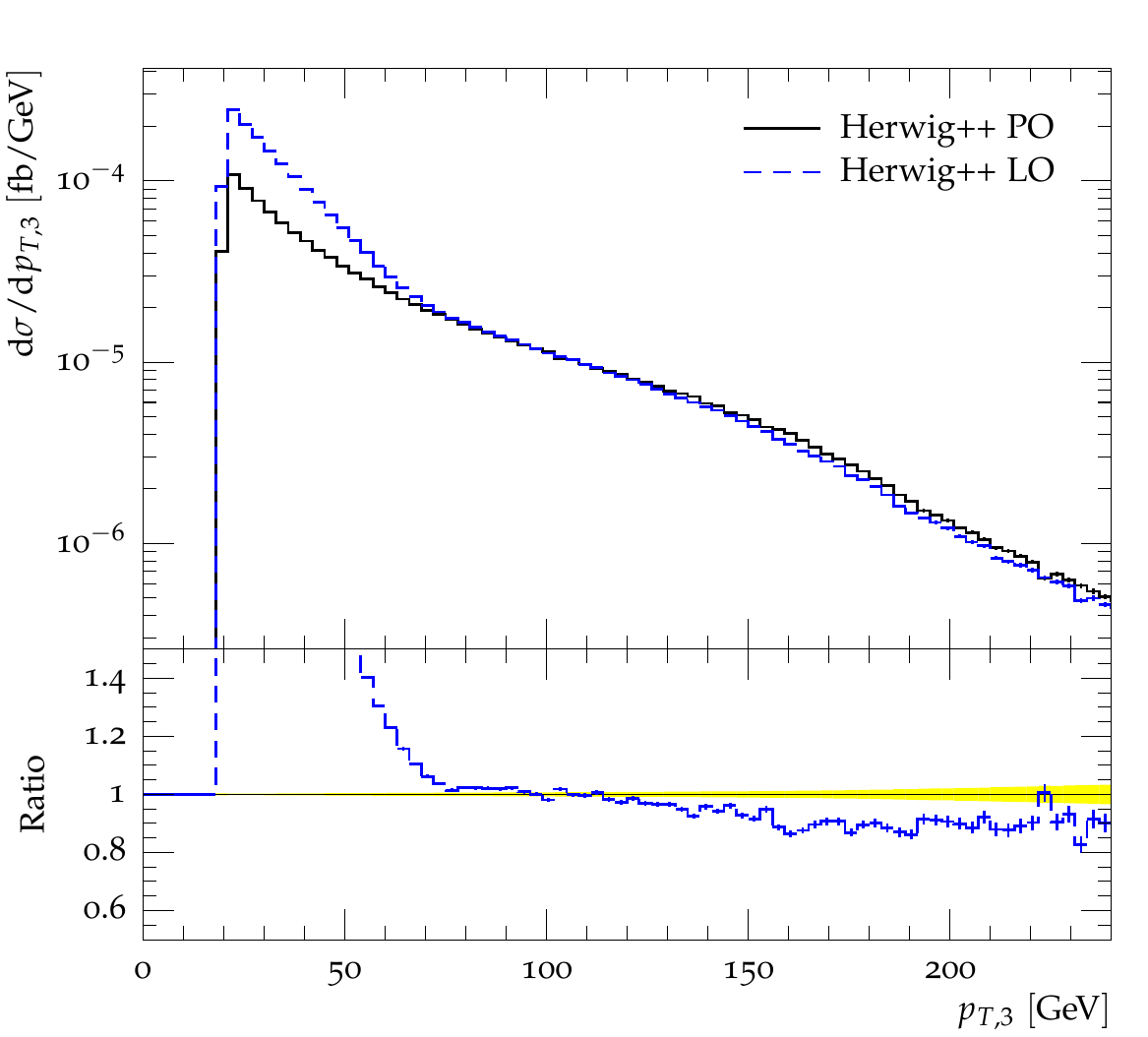}

  \caption{Comparison of parton level distributions generated with and without the POWHEG correction for the decay \(\tilde{g} \rightarrow  \tilde{t}_1 \, \bar{t}\) with stable decay products.  Results are for the CMSSM model with  \(m_0 = 1220 \GeV\), \mbox{\(m_{1/2} = 660 \GeV\)}, \(\tan{\beta} = 10\), \(A_0 = 0\) and \(\mu >0\) and LHC events with \(\sqrt{s}=8 \TeV\).  Shown are the \(p_T\) distributions of the the third hardest jet in the rest frame of the \(\tilde{g}\).  }
  \label{fig:gluino stable}

\end{figure}

The transverse momentum distribution of the third hardest jet in the rest frame of the \(\tilde{g}\) were also plotted and are shown in Fig.~\ref{fig:gluino stable}.  To focus on the effect of the POWHEG correction, the decay products, \(\bar{t}\) and \(\tilde{t}_1 \), were not allowed to decay further.  The blue dashed and black solid lines in Fig.~\ref{fig:gluino stable} correspond to the parton shower and POWHEG correction distributions respectively.  The bottom panel of the plot shows the ratio of the parton shower and POWHEG corrected results and in both distributions error bars are included to indicate statistical uncertainty.  As in Sect.~\ref{sec:chiResults}, we find that the POWHEG correction decreases the total number of events passing the jet \(p_T\) selection criterion.  The effect is more pronounced in this case, with an \(\mathcal{O} \left(40 \% \right )\) reduction. The parton shower distribution significantly exceeds the POWHEG corrected distribution at small \(p_{T,3}\), however, at higher values of \(p_{T,3}\) the two distributions are similar.  At lower values of \(p_{T,3}\), the main contribution to the third hardest jet in the POWHEG corrected distribution is from the hardest emission in the decay, generated using the real-emission matrix element.  Therefore, we expect the uncorrected distribution to exceed the corrected one in this region.  However, the maximum possible \(p_T\) of the gluon generated 

\noindent by the POWHEG correction is\footnote{The value of \(p^{\rm{max}}_T\) was calculated using the formula for \(p^{\rm{max}}_T\) in top quark decay, given on page~\pageref{eq:SudakovOver},  with the replacements \(m_t \rightarrow m_{\tilde{g}}\), \(m_W \rightarrow m_{\tilde{t}_1}\) and \(m_b \rightarrow m_t\) } \(p^{\rm{max}}_T \approx 75\GeV\).  Jets contributing to the POWHEG corrected distribution above this limit include a number of other partons generated by the normal parton shower in addition to the hardest emission.  This reduces the effect of the correction at higher values of \(p_{T,3}\).  Therefore, we find that applying the POWHEG correction has a more significant impact on the number of events passing selection criteria when the value of the \(p_{T,3}\) selection criterion lies below \(p^{\rm{max}}_T\) of the gluon produced in the POWHEG correction.

\section{Conclusions} 
In this work, we used the real-emission matrix element to generate hard QCD radiation in a range of particle decays in the \Herwig\ event generator.  This method is particularly relevant to new physics searches based on the decays of heavy new particles.  The POWHEG corrections to these decays can change the shapes of certain experimental observables, thus altering the number of signal events passing selection criteria and modifying the exclusion bounds that can be set on the masses of the new particles.  This correction will be available in \Herwig\ version 2.7.

The algorithm used to implement the POWHEG style correction in \Herwig\ was described in detail for the decay \(t \rightarrow W b\).  Dalitz style distributions of the first emission in the conventional parton shower and POWHEG corrected approach were produced and showed that, while the POWHEG style correction ensures the majority of emissions lie in the soft and collinear limits, the parton shower has erroneous, unphysical regions of high emission density.  This causes the parton shower to overpopulate the high \(p_T\) regions of phase space.  Differential distributions of the minimum jet separation and logarithm of the jet measure were also generated with the POWHEG style correction and compared to those generated with the existing \Herwig\ implementation of hard and soft matrix element corrections.  The two techniques exhibit a high level of agreement therefore demonstrating the validity of our approach.  In addition to this, distributions were generated using the normal parton shower.  In agreement with the results from the Dalitz plots, these distributions were found to be considerably harder than those generated with the matrix element or POWHEG style corrections.

The impact of applying the POWHEG style correction to a BSM decay was studied by plotting the invariant mass distribution of dijets produced in the decay of the lightest RS graviton, \(G \rightarrow gg\) and \(G \rightarrow q \bar{q}\).  Applying the POWHEG correction was found to have a considerable impact on the height of the distribution in the dijet mass peak.  The number of events passing selection criteria in the mass range \(2.1 \TeV \leq m_{jj} \leq 2.3 \TeV \) dropped by \(\mathcal{O} \left(40 \% \right) \) when the correction was applied.  This is a consequence of the dijet invariant mass not including the hardest emission in the shower that carries a significant fraction of the graviton's momentum when it is simulated using the real-emission matrix element.  The sizable impact of the correction in this scenario illustrates the importance of including higher order corrections when optimising experimental searches.  

The impact of the POWHEG correction was also investigate for two decays in the CMSSM model by studying the transverse momentum distributions of the hardest jet generated by the shower.   At values of the transverse momentum less than the upper limit of the POWHEG correction, it was found that the POWHEG corrected distributions were significantly reduced with respect to those generated with the conventional parton shower.  Above this cutoff, the normal parton shower and POWHEG corrected distributions were found to be similar.  

In this work, we have used the POWHEG formalism to improve the simulation of hard radiation in particle decays and studied the resulting effect on a number of distributions.  However, hard radiation in the initial-state parton shower can also have a significant impact on these distributions.  Hence, in order to achieve accurate simulation of hard radiation in BSM processes we must also include effects from the initial-state shower.  Using the POWHEG formalism to improve the simulation of the hardest initial-state emission in the shower will be the subject of future work. 

\addtocontents{toc}{\protect\setcounter{tocdepth}{-1}}
\section{Acknowledgements}
We are grateful for help from the other members of the \Herwig\ collaboration.  This work was supported by the Science and Technology Facilities Council.  We also acknowledge the support of the European Union via MCNet.

\bibliography{Herwig}
\end{document}